# A Survey on Web Multimedia Mining

Pravin M. Kamde[1], Dr. Siddu. P. Algur[2]

[1] Assistant Professor, Department of Computer Science & Engineering, Sinhgad College of Engineering, Pune, Maharashtra,
Email: pravinkamde@rediffmail.com

[2] Professor & Head, Department of Information Science & Engineering, BVB College of Engineering & Technology, Hubli, Karnataka,
Email: algursp@bvb.edu

## Abstract

*Modern developments in digital media technologies has made transmitting and storing large amounts of multi/rich media data (e.g. text, images, music, video and their combination) more feasible and affordable than ever before. However, the state of the art techniques to process, mining and manage those rich media are still in their infancy. Advances developments in multimedia acquisition and storage technology the rapid progress has led to the fast growing incredible amount of data stored in databases. Useful information to users can be revealed if these multimedia files are analyzed. Multimedia mining deals with the extraction of implicit knowledge, multimedia data relationships, or other patterns not explicitly stored in multimedia files. Also in retrieval, indexing and classification of multimedia data with efficient information fusion of the different modalities is essential for the system's overall performance. The purpose of this paper is to provide a systematic overview of multimedia mining. This article is also represents the issues in the application process component for multimedia mining followed by the multimedia mining models.*

## 1. Introduction

Now days the World Wide Web is a popular and interactive medium to discriminate information. The web is huge, diverse, and dynamic and thus raises the scalability, multimedia data and temporal issues respectively. The need to understand large, complex, information-rich data sets is common to virtually all fields of business, science, and engineering. The ability to extract useful knowledge hidden in these data and to act on that knowledge is becoming increasingly important in today's competitive world. The entire process of applying a computer-based methodology for discovering and extracting knowledge from web documents is a web mining. Web multimedia mining is inherently at the cross road of research from several multi-discipline like computer vision, multimedia processing, multimedia retrieval, data mining, machine learning, database and artificial intelligence.

A multimedia database system includes a multimedia database management system (MM-DBMS), which manages and also provides support for storing, manipulating, and retrieving multimedia data from a multimedia database, a large collection of multimedia objects, such as image, video, audio, and hypertext data, is discuss in [1].

A large number of techniques have been proposed ranging from simple measures (e.g. color histogram for image, energy estimates for audio signal) to more sophisticated systems like automatic summarization of TV programs [2], speaker emotion recognition from audio [3] in the area of multimedia mining system that extract semantic about the web multimedia.

The following sections brief about the web mining. Sections 3 describes mining multimedia databases, 3.2, 3.3, 3.4 and 3.5 will discuss text, image, video, and audio mining respectively. Section 4 describes the application process and reprocessing that includes data cleaning, normalization, transformation and feature selection in multimedia mining. In Section 5 supervised and unsupervised models are discuses.

## 2. WEB MINING TAXONOMY

Three distinct categories based on [4] and [5], are: the application of data mining techniques to extract and prepare knowledge from Web content (include text, image and video), structure (hyperlinks between documents), and usage (logs of web sites). Figure 1 shows web mining taxonomy, according to the kinds of data to be mined:

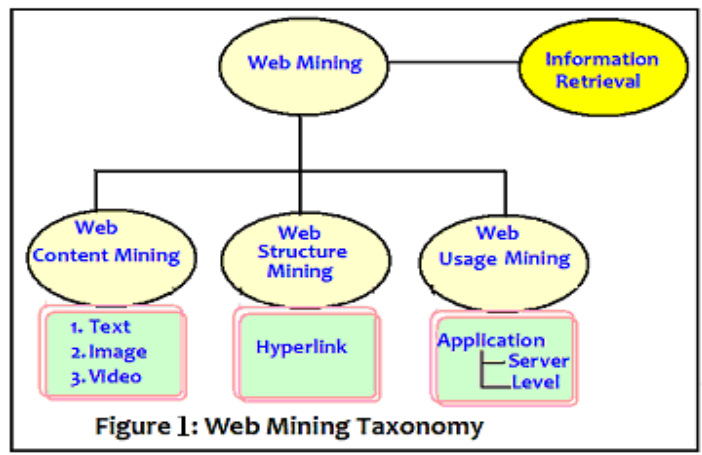

Figure 1: Web Mining Taxonomy

### 2.1. Web Content Mining

Web Content Mining is the process of extracting useful information from the contents of Web documents. It may consist of text, images, audio, video information which is used to convey to the users about that documents. Text mining and its application to Web content has been the most widely researched. Some of the research issues addressed in text mining are, topic discovery, extracting association patterns, clustering of web documents and classification of Web Pages. Web content mining issues in term of Information Retrieval (IR) and Database (DB) view verses data representation, method and application categories is discuss and summarised in [6]. While extracting the knowledge from images - in the fields of image processing and computer vision - the application of these techniques to Web content mining has not been very rapid.

### 2.2. Web Structure Mining

The structure of a typical Web graph consists of Web pages as nodes, and hyperlinks as edges connecting between two related pages it is the process of discovering structure information from the Web.
_ Hyperlinks: A Hyperlink is a structural unit that connects a Web page to different location, either with in (intra-Document Hyperlink) and/or another Web page (inter-Document Hyperlink)
_ Document Structure: the content within a Web page can also be organized in a tree-structured format, based on the various HTML and XML tags within the page.

Some algorithms proposed to model the web topology such as HITS, Page-rank, improved in HITS by outer filtering, web page categorization address in [6].

## 2.3. Web Usage Mining

Web Usage Mining is the application of data mining techniques to discover interesting usage patterns from web data, in order to understand and better serve the needs of web-based applications. Some of typical data mining methods used to mine the usage data after the data have been pre-processed in desired form. Modification of typical data mining is used composite associative rule, extraction of traditional sequence discovery algorithm, web usage using graph representation, is discussed in and can referred from [6].

## 3. MINING MULTIMEDIA DATABASES

Multimedia has been the major focus for many researchers around the world. Many techniques for representing, storing, indexing, and retrieving multimedia data have been proposed. Most of the studies done are confined to the data filtering step of the KDD process. In [7], Czyzewski demonstrated how KDD methods can be used to analyze audio data and remove noise from old recordings. Chien et al. in [8] use knowledge based AI techniques to assist image processing in a large image database generated from the Galileo mission.

A multimedia data mining system prototype, MultiMediaMiner- includes the construction of a multimedia data cube which facilitates multiple dimensional analyses of multimedia data, primarily based on visual content, and the mining of multiple kinds of knowledge, including summarization, comparison, classification, association, and clustering [9].

Thus, in multimedia documents, knowledge discovery deals with non-structured information. In general, the multimedia files from a database must be first pre-processed to improve their quality followed by feature extraction. With the help of generated features, information models can be devise using data mining techniques to discover significant patterns as shown in figure 2. Multimedia data mining refers to pattern discovery, rule extraction and knowledge acquisition from multimedia database, as discussed in [10].

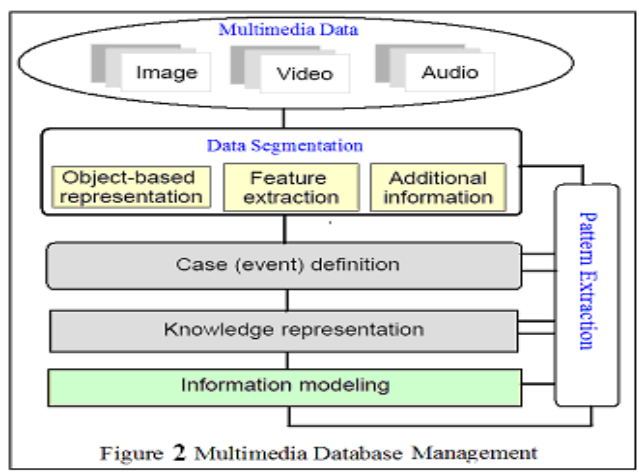

Figure 2 Multimedia Database Management

## 3.1. Overview

Multimedia is harder to fit into typically data mining models. Image and video of different entities have some similarity - each represents a view of a building - but without clear structure such as "these are pictures of the front of buildings" it is difficult to relate multimedia mining to traditional data mining. Multimedia generally gives a lot of data on each entity, but not the same data for each entity.

A second difference between multimedia mining and structured data mining is the sequence or time element. Multimedia often captures an entity changing over time. Video and audio are clearly ordered, and even text has little meaning without sequence. Time series mining analyzes the change to one or more values over time. Multimedia is more complex - as the sequence progresses, the concept being represented may change as well. This is obvious with video, where a camera may slate or objects in the scene may move. Understanding and representing changes in the mining process is necessary to mine multimedia data [11].

These heterogeneous databases could be first integrated and then mined or one could apply mining tools on the individual databases and then combine the results of the various data miners are carried out via the Multimedia Distributed Processor (MDP). If the data is to be mined first, the data miner augments the corresponding MM-DBMS and the results of the data miners are integrated via the MDPs as shown in figure 3. Since there is much to be done on mining individual data types such as text, images, video, and audio, mining combinations of data types is still a challenge [11].

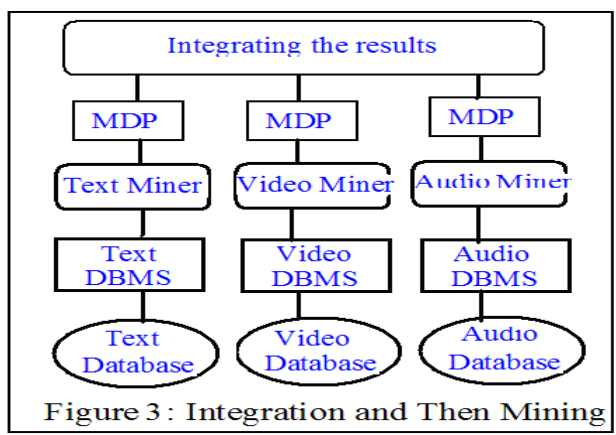

Figure 3: Integration and Then Mining

## 3.2. Text Mining

Current research in the area of text mining tackle the problems of text representation, classification, clustering, information extraction or the search for and modelling of hidden patterns. In this context the selection of characteristics, domain knowledge and domain-specific procedures play an important role.

Information retrieval systems and text processing systems have been developed are quite sophisticated and can retrieve documents by specifying attributes or key words. However, in order to be able to define at least the importance of a word within a given document, usually a vector representation is used, where for each word a numerical "importance" value is stored. The predominant approaches based on this idea are the vector space model [35], the probabilistic model [36] and the logical model [37].

Text mining or text data mining, the process of finding useful or interesting patterns, models, directions, trends, or rule from unstructured text, is used to describe the application of data mining techniques to automated discovery of knowledge from text [38]. Text mining has been viewed as a natural extension of data mining [39], sometimes considered as a task of applying a same data mining techniques to specific domain [40]. This reflect the fact that advent of text mining relies on the burgeoning field of data mining to a great degree.

Text categorization is a conventional classification problem applied to the textual domain. It solves the problem of assigning text content to predefined categories. In the learning stage, the labeled training data are first pre-processed to remove unwanted details and to "normalize" the data [13]. The keyword extraction from the document is identifying summarize the contents of the document. The common English removed using an "ignore-list" of words during the pre-processing stage. And a good heuristic is applied for words that occur frequently in documents of the same class.

## 3.3 Image Mining

Image mining is the concept used to detect unusual patterns and extract implicit and useful data from images stored in the large data bases. Therefore, we can say that image mining deals with making associations between different images from large image databases. Image mining is used in variety of fields like medical diagnosis, space research, remote sensing, agriculture, industries, and also handling hyper spectral images. Images include maps, geological structures, and biological structures and even in the educational field, explained in [12].

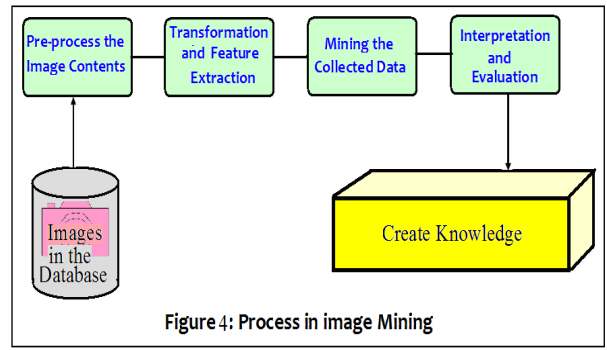

Figure 4: Process in image Mining

The fundamental challenge in image mining is to reveal out how low-level pixel representation enclosed in a raw image or image sequence can be processed to recognize high-level image objects and relationships. Ji Zhang, Wynne Hsu and Mong Li Lee [41] proposed an efficient information-driven framework for image mining. In that they made out four levels of information: Pixel Level, Object Level, Semantic Concept Level, and Pattern and Knowledge Level. To achieve that High-dimensional indexing schemes and retrieval techniques are incorporated in the framework to maintain the flow of information among the levels. Ji Zhang, Wynne Hsu and Mong Li Lee [42] highlighted the need for image mining in the era of rapidly growing amounts of image data and pointed out about the unique characteristics of image databases. In addition, it is also examined function-driven and information-driven frameworks for image mining. The image may still have other dimensions. Selecting a subset of features is a method for reducing the problem size is stated in [21].

### 3.3.1. Feature Extraction from Color Images

Image categorization classifies images into semantic databases that are manually recategorized. In the same semantic databases, images may have large variations with dissimilar visual descriptions (e.g. images of persons, images of industries etc.). In [14], the authors distinguish three types of feature vectors for image description: 1) pixel level features, 2) region level features, and 3) tile level features.

Moreover, researchers proposed in [15] an information-driven framework that aims to highlight the role of information at various levels of representation, which adds one more level of information: the Pattern and Knowledge Level that integrates domain, related alphanumeric data and the semantic relationships discovered from the image data.

Keiji Yanai [43] describes a generic image classification system with an automatic knowledge acquisition mechanism from the Web. Nick Morsillo, Chris Pal, Randal Nelson [44] presented a technique that allows a user to reduce noisy search results and characterize a more precise visual object class. This approach is based on semi-supervised machine learning in a novel probabilistic graphical model made of both generative and discriminative elements. Bingbing Ni, Zheng Song, Shuicheng Yan [45] presented an automatic image mining system in web which builds a universal human age estimator based on facial information, which can be used to all racial groups and various image qualities. The relations between different modalities of Web images could be very useful for Web image retrieval. Ruhan He Wei Zhan [46] investigates the multi-modal associations between two basic modalities of Web images, i.e. keyword and visual feature clusters, by multi-model association rule. Automatic image classification is a demanding research topic in Web image mining. Rong Zhu, Min Yao and Yiming Liu [47] formulated image classification problem as the calculation of the distance measure between training manifold and test manifold.

Framework for mining images by colour content is proposed in [51], framework provides the possibility of use 5 distance function for evaluation of similarity among images and 2 type of quantization. A recursive HSV-space segmentation technique to identify perceptually prominent color areas discussed in [52]. The average color vector of these extracted areas is then used to build the image indices, requiring very little storage. A framework focuses on color as feature using Color Moment and Block Truncation Coding (BTC) to extract features for image dataset is proposed in [53]. Then K-Means clustering algorithm is conducted to group the image dataset into various clusters. In [54] proposed an application of Binary Thresholded Histogram (BTH), a color feature description method, to the creation of a metadatabase index of multiple image databases. Paper [55] described about an Image mining techniques which is based on the Color Histogram, texture of that Image. The query image is taken then the Color Histogram and Texture is produced and based on this the resultant Image is found. They have investigated a histogram-based search methods and color texture methods in two different color spaces, RGB and HSV. Histogram search differentiate an image by its color distribution. It is shown that images retrieved by using the global color histogram may not be semantically associated even though they share similar color distribution in some results. There is need to conduct more research on image mining to see if data mining techniques could be used to classify, cluster, and associate images.

## 3.4. Video Mining

Mining video data is even more complicated than mining image data. One can regard video to be a collection of moving images, much like animation. The important areas include developing query and retrieval techniques for video databases, including video indexing, query languages, and optimization strategies.

In video mining, there are three types of videos: a) the produced (e.g. movies, news videos, and dramas), b) the raw (e.g. traffic videos, surveillance videos etc), and c) the medical video (e.g. ultra sound videos including echocardiogram). Higher-level information from video includes: i) detecting trigger events (e.g. any vehicles entering a particular area, people exiting or entering a particular building), ii) determining typical and anomalous patterns of activity, generating person-centric or object-centric views of an activity, and iii) classifying activities into named categories (e.g. walking, riding a bicycle), clustering and determining interactions between entities [18]. Shot detection methods can be classified into many categories: pixel based, statistics based, transform based, feature based and histogram based, discussed in [19]. To segment video, color histograms, as well as motion and texture features can be used as describe in [20].

For example, one could examine video clips and find associations between different clips. Other one could find unusual patterns in video clips. Pattern matching in video databases possible when one has predefined images and then matches these images with the multiple video clips and analyzed a video clips.

Past work in completely unsupervised structure discovery includes that of Xie et al [48], Foote et al [49] and Peker [50]. Xie et al use an unsupervised Hierarchical Hidden Markov Model (HHMM) framework to discover patterns in soccer video. They use low-level features such as motion intensity and dominant color at the lower level of the HHMM and binary labels at the upper level of the HHMM.

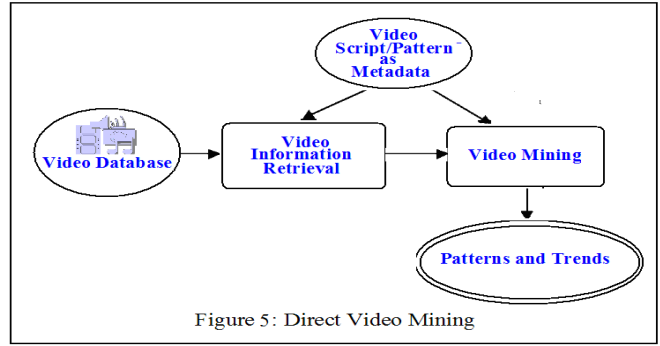

Figure 5: Direct Video Mining

In general, the purpose has been to identify video segments for a human to watch - the actionable knowledge is generated by the person watching the video. The next step requires understanding what types of knowledge we would hope to gain from video mining.

### 3.5. Audio Mining

Since audio is a continuous media type like video, the techniques for audio information processing and mining are similar to video information retrieval and mining. Audio data could be in the form of radio, speech, or spoken language. Even television news has audio data, and in this case audio may have to be integrated with video and possibly text to capture the annotations and captions. To mine audio data, one could convert it into text using speech transcription techniques. Audio data could also be mined directly by using audio information processing techniques and then mining selected audio data.

The researchers used perceptual features such as loudness, brightness, pitch etc, to process on audio extracted as describe in [16]. However, most frequently used features for audio classification are discussed in [17].

In general, audio mining (as opposed to mining transcribed speech) is even more primitive than video mining. While a few papers have appeared on text mining and even fewer on images and video mining, work on audio mining is just beginning.

## 4. PROCESS OF APPLICATION OF MULTIMEDIA MINING

The model of applying multimedia mining in different multimedia types is present in figure 6. Data collection is the starting point of a learning system, as the quality of raw data determines the overall achievable performance. Then, the goal of data pre-processing is to discover important features from raw data.

Data pre-processing includes data cleaning, normalization, transformation, feature selection, etc. Learning can be straightforward, if informative features can be identified at pre-processing stage.

Detailed procedure depends highly on the nature of raw data and problem's domain. The product of data pre-processing is the training set. Given a training set, a learning model has to be chosen to learn from it and make multimedia mining model more iterative.

Higher complexity found on compared data mining with multimedia mining: a) the huge volume of data, b) the variability and heterogeneity of the multimedia data (e.g. diversity of sensors, time or conditions of acquisition etc) and c) the multimedia content's meaning is subjective. Application and system of multimedia data mining based on process discussed is surveyed in following sub-section.

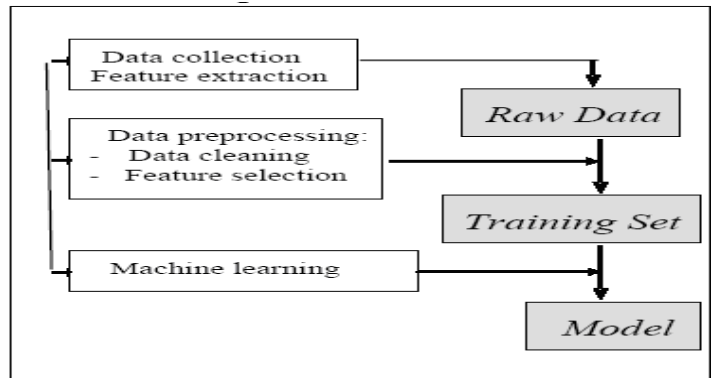

Figure 6 Multimedia mining process

## 4.1. Applications and Systems

Satellite data is used in many different areas ranging from agriculture, forestry, and environmental studies. The applications using satellite data include measurements of crop and timber acreage, forecasting crop yields and forest harvest, monitoring urban growth, mapping of ice for shipping, mapping of pollution, recognition of certain rock types, and many others.

For example, the CONQUEST system [30] combines satellite data with geophysical data to discover patterns in global climate change. The SKICAT system [31] integrates techniques for image processing and data classification in order to identify 'sky objects' captured in a very large satellite picture set. An example of video and audio data mining can be found in the Mining Cinematic Knowledge project [32], which created a movie mining system by examining the suitability of existing concepts in data mining to multimedia.

Moreover, the analysis and mining of traffic video sequences in order to discover information (such as vehicle identification, traffic flow, and the spatio-temporal relations of the vehicles at intersections) provide an economic approach for daily traffic operations. There are some multimedia data mining frameworks [33] for traffic monitoring systems. Furthermore, various methods for the detection of faces in images and image sequences are reported in [34]. Detection of generic sport video documents seems almost impossible due to the large variety in sports.

## 5. MODELS FOR MULTIMEDIA MINING

Multimedia classification and clustering based on the supervised and unsupervised learning.

### 5.1 Classification models

Machine learning (ML) and meaningful information extraction can only be realized. An overview of existing supervised models works in decision trees is provided in [22]. Decision trees can be translated into a set of rules by creating a separate rule for each path from the root to a leaf in the tree. The rules can also be directly induced from training data using a variety of rule-based algorithms [23]. Artificial Neural Networks (ANNs) are another method of inductive learning, based on computational models of biological neurons and networks [24]. The Support Vector Machines (SVMs) is the newest technique that considers the notion of a "margin". Maximising the margin and thereby creating the largest possible distance between the separating hyperplane and the instances on either side of it, is proven to reduce an upper bound on the expected generalisation error [25].

### 5.2 Clustering Models

In unsupervised classification, the problem is to group a given collection of unlabeled multimedia files into meaningful clusters according to the multimedia content without a priori knowledge. Clustering algorithms can be categorized into partitioning methods, hierarchical methods,

The International Journal of Multimedia & Its Applications (IJMA) Vol.3, No.3, August 2011

density-based methods, grid-based methods, and model-based methods. An excellent survey of clustering techniques can be found in [26].

Density-based clustering algorithms try to find clusters based on density of data points in a region. The key idea of density-based clustering is that, for each instance of a cluster, the neighbourhood of a given radius has to contain at least a minimum number of instances [27]. Grid-based clustering algorithms first quantize the clustering space into a finite number of cells (hyper-rectangles) and then perform the required operations on the quantized space.

### 5.3 Association Rules

Rule based classification is based on the occurrences similarity of pattern or entity in the specific domain data set. There are three measures of the association: support, confidence and interest. The support factor indicates the relative occurrences of both X and Y within the overall data set of transactions. It is defined as the ratio of the number of instances satisfying both X and Y over the total number of instances. The confidence factor is the probability of Y given X and is defined as the ratio of the number of instances satisfying both X and Y over the number of instances satisfying X. The support factor indicates the frequencies of the occurring patterns in the rule, and the confidence factor denotes the strength of implication of the rule. The interest factor is a measure of human interest in the rule.

Relatively little research has been conducted on mining multimedia data [28]. There are different types of associations: association between image content and non image content features. Association mining in multimedia data can be transformed into problems of association mining in traditional transactional databases. Therefore, mining the frequently occurring patterns among different images becomes mining the frequent patterns in a set of transactions. In [29], the authors extend the concept of content-based multimedia association rules using feature localization. They introduced the concept of progressive refinement in discovery of patterns in images.

## 6. CONCLUSION

Systematic reviews of existing works reveal gaps and opportunities in the field of multimedia mining, which contents non-structured heterogeneous information: audio, video, image, speech, text, graphics, icons, web logs, etc. The multimedia mining, knowledge extraction plays crucial role in multimedia knowledge discovery.

We have addressed the overview and use of multimedia database management systems. And discuss on mining for multimedia data, process of application of multimedia mining followed by models for multimedia mining classification and issues.

Issues in multimedia mining: too much is lost when the sequence of multimedia is ignored. The two approaches to mining ordered data, time series and event sequences, are may not be adequate for in certain cases.

In text mining there are two open problems: polysemy, synonymy. Polysemy refers to the fact that a word can have multiple meanings. Synonymy means that different words can have the same/similar meaning.

In image mining an open problem remains: the combination of different types of image data. In audio and video mining, a fundamental open problem also remains: The combination of information across multiple media (combining video and audio information into one comprehensive score).

An interesting research direction on web content mining is the integration of heterogeneous information sources.

The International Journal of Multimedia & Its Applications (IJMA) Vol.3, No.3, August 2011

**Authors**

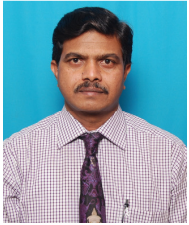

**Pravin M. Kamde**, is working as a Assistant Professor in the Department of Computer Engineering, at Sinhgad College of Engineering, Pune. He has received degree of B.E. (Computer Science and Engineering) from SGGS College of Engineering and Technology, Nanded, Marathwada University Auragabad, in 1993, M.E.(Computer Science and Engineering) in 2004 from Walchand College of Engineering, Sangli, Shivaji University. He has 2 International journal publications, 5 International conferences, and 12 National conferences publication. Currently he is India. His research interests include Content-based Image and Video Retrieval, Web Multimedia Mining, and Image Processing

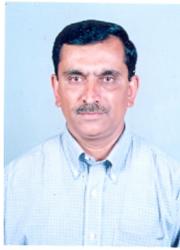

**Dr. Siddu P. Algur, is** working as a Professor and Head, Dept. of Information Science and Engineering at BVB College of Engineering and Technology, Hubli. He received B.E. degree in Electrical and Electronics from Mysore University, Karnataka, India, in 1986. He received his M.E. degree in Information Science and Engineering from NIT, Allahabad, India, in 1991.He obtained Ph. D. degree from the Department of P.G. Studies and Research in Computer Science at Gulbarga University, Gulbarga. Mr. Algur worked as Lecturer in the department of E & E at KLE Society's College of Engineering and Technology from September 1986 to October 1992. From October 1992 to August 2007 he worked as Lecturer and Assistant Professor in the department of Computer Science and Engineering at SDM College of Engineering and Technology, Dharwad. His research interest includes Data Mining, Web Mining, and Information Retrieval from the web and Knowledge discovery techniques. He has published 14 research papers in peer reviewed International Journals and chaired the sessions in many International conferences. Tel (off): +91 836 2378402, e-mail: algursp@bvb.edu.